\documentclass[11pt]{article}
\usepackage[dvips]{epsfig}
\begin{document}                

\begin{center}
{\bf Tidal Theory of the Thermal Wind\\}
\vspace{3.mm}
{Detlev M\"{u}ller\\}
MPIMET\\
{\it D-20146 Hamburg, Germany\\}
\end{center}

\begin{abstract} 
The baroclinic instability problem is considered in the framework of Laplacian tidal 
theory. The Hilbert space of the quasigeostrophic vorticity budget is spanned by 
spheroidal functions. The fluid is linearly stable against quasigeostrophic disturbances. 
As the essential source of irregular ocean-atmosphere motions, baroclinic instability is 
ruled out by tidal theory. The midlatitude $\beta$-plane budget of vorticity fluxes is 
inconsistent with basic laws of motion on the rotating spherical surface. Realistic 
numerical simulations of global wave dynamics and dynamical circulation instabilities 
require a covariant account of fluid motion on the spherical planet.
\end{abstract}

\vspace{5.mm}
\noindent
Pacs numbers: 47.32.-y, 47.35.+i, 92.10.Hm, 92.60.Dj

\vspace{5.mm}
\noindent
{\bf I. Introduction}

Major difficulties in numerical climate simulations result from the fact that the potential
energy in the ocean-atmosphere system is closely associated with large-scale features of its
density field while a considerable fraction of its kinetic energy resides on fairly small
eddy scales \cite{1}. State-of-the-art models for weather prediction and climate simulation 
capture large-scale features of the global circulation with some degree of realism if they 
are determined by large-scale features of topography and external forcing. However, simulation 
of the dynamical control of density by transfer processes between the small scales of kinetic 
energy and the large potential energy scales is less satisfactory \cite{2}. The key to these 
energy- and vorticity-fluxes is fluid instability. Of the numerous instabilities in the 
climate system none is considered as fundamental as baroclinic instability \cite{3}. This 
concept refers to the growth of weakly divergent Rossby waves in a stably stratified 
and vertically sheared fluid on the rotating spherical surface. The contemporary understanding 
of cyclogenesis, predictability-limits and the transition to chaos,  turbulence and 
stochasticity in the climate system revolves essentially around this process. Hence, it also
provides the paradigm for the design and interpretation of irregular fluid motion in numerical 
circulation models, ranging from resolved scales down to the parameterization of subscale 
processes.

Baroclinic instability theory invokes four sets of approximations. First: shallow water 
theory with constant mean layer-thicknesses and -velocities \cite{3}. Second: quasigeostrophy. 
For tidal theory, Longuet-Higgins \cite{4} has shown that this approximate closure of the 
vorticity budget yields a meaningful and elegant Rossby wave filter of Laplace's tidal 
equation. In baroclinic fluids this approach governs the thermal wind. Third: the midlatitude 
$\beta$-plane. To avoid formal difficulties with spherical coordinates, a tangential cartesian 
plane is pinned to the sphere at some extratropical latitude in analogy to Kelvin's 
well-established equatorial $\beta$-plane. Fourth: a number of scaling assumptions which neglect 
fluid velocities relative to planet rotation and stratification relative to barotropicity 
wherever it appears uncritical. Quantitatively, these assumptions seem generally to be 
justified \cite{5}.

The paper at hand considers the baroclinic instability problem of Laplace's tidal theory, 
independent of the midlatitude $\beta$-plane and scaling assumptions. Tidal theory differs in
two aspects from current circulation theories and models: it is Newtonian (namely: covariant) 
and accounts consistently for the globe's sphericity \cite{6}. The relationship of its 
analytical structure and the theory of Heun functions \cite{7} is increasingly understood. 
While a transformation of the tidal equation into Heun's equation is not known, general 
approximations and exact special cases can be expressed in terms of Heun functions. These 
concepts provide a sound  foundation for Rayleigh stability analysis in the framework of 
global wave-circulation theory. On the basis of covariant fluid dynamics on the rotating 
spherical surface it will be shown that Rossby wave growth in baroclinic fluids plays a
far lesser role for irregular ocean-atmosphere motions than currently thought. 

The geometrically and dynamically consistent way to avoid formal difficulties with spherical
coordinates is the use of index notation and covariant differentiation. Here, indices 
$m,n,\ldots=1,2$ run over longitude $\lambda$ and latitude $\varphi$ while braced indices 
$(\ell)=1,2$ refer to the top and bottom layer of the fluid and are not subject to the 
summation convention. Covariant differentiation will be denoted by a semicolon. For details 
of the notation see \cite{6}. In this formalism, equations in curvilinear coordinates look 
widely similar to corresponding equations in cartesian coordinates with geometrical details
consistently absorbed into core symbols and indices. Thus, formulas emphasize the physical 
structure of the problem.

\vspace{5.mm}
\noindent
{\bf II. Tidal Equations}

The problem is considered in terms of the bishallow water equations on the rotating spherical 
surface \cite{6}. Rayleigh stability theory requires the linearization of such equations 
around the considered basic state. For both layers, the mean layer mass per unit area and
hence the concentration are here assumed to be constant 
\[R=R_{(1)}+R_{(2)}= const, \hspace{10.mm}  r=R_{(1)}/R=const.\]
Thus, the effective pressure $P(R,r)$ and the interfacial potential $\mu(R,r)$ 
\[P=\frac{1}{2}\,\gamma_{(2)}\,(1+\delta r^{2})\,R^{2}, \hspace{10.mm} \mu=
\gamma_{(2)}\,\delta\, Rr,\]
are also constant where $\gamma_{(\ell)}=g/\rho_{(\ell)}$, $g$ the gravitational acceleration 
and  
\[\delta=(\rho_{(2)}-\rho_{(1)})/\rho_{(1)}>0\] 
the positive definite stratification parameter. The spherical generalization of a constant 
mean barycentric velocity $V_{n}$ and a constant mean vertical shear $W_{n}$ are given by 
\[V_{n}=a^{2}U_{0}(cos^{2}\varphi,0), \hspace{10.mm} W_{n}=a^{2}W(cos^{2}\varphi,0)\]
with Earth's radius $a$ and constant angular velocities
\[U_{0}=rU_{(1)}+(1-r)U_{(2)},\hspace{5.mm} U_{1}=(1-r)U_{(1)}+rU_{(2)},\hspace{5.mm} 
W=U_{(1)}-U_{(2)}.\]
The barycentric and baroclinic potential vorticities of this circulation , $Z_{0}$ and $Z_{1}$,
are obtained as 
\[RZ_{0}=F_{0}=2(\Omega+U_{0})sin\varphi, \hspace{10.mm} RZ_{1}=F_{1}=
2(\Omega+U_{1})sin\varphi\]
while $S=Wsin\varphi$. This circulation is driven by an external, meridionally varying 
surface pressure and the value of the vertical shear $W$ is determined by $\Omega$, the 
stratification parameter $\delta$ and the equator-to-pole gradient of the surface 
pressure \cite{6}. For small-amplitude perturbations $(m,\eta,j_{n},i_{n})$ of the state
vector: layer mass, concentration, barycentric and baroclinic mass flux, linearization of the 
bishallow water equations around this circulation leads to the tidal problem \cite{6}
\begin{equation}
d_{0}m+j^{n};_{n}=-RW^{n}\partial_{n}\eta
\end{equation}
\begin{equation}
Rd_{1}\eta+i^{n};_{n}=-r_{12}W^{n}\partial_{n}m
\end{equation}
\begin{equation}
d_{0}j_{n}+\epsilon_{mn}F_{0}j^{m}+\partial_{n}p_{\ast}
=-W^{m}\partial_{m}i_{n}-2\epsilon_{mn}Si^{m}
\end{equation}
\begin{equation}
d_{1}i_{n}+\epsilon_{mn}F_{1}i^{m}+\partial_{n}\mu_{\ast}=
-r_{12}(W^{m}\partial_{m}j_{n}+2\epsilon_{mn}Sj^{m})
\end{equation} 
where $r_{12}=r(1-r)$ and $d_{0/1}=\partial_{t}+U_{0/1}\partial_{\lambda}$. The linearized 
pressure emerges as
\[p_{\ast}=(\partial_{R}\,P)_{r}\,m+(\partial_{r}\,P)_{R}\,\eta=c^{2}p_{1}m+c^{2}p_{2}R\eta\]
while one finds for the linearized interfacial potential
\[\mu_{\ast}=r_{12}R(\partial_{R}\,\mu)_{r}\,m+r_{12}R(\partial_{r}\,\mu)_{R}\,\eta=
c^{2}\mu_{1}m+c^{2}\mu_{2}R\eta.\]
Here, $c^{2}=\gamma_{(2)}R$ and 
\[p_{1}=1+\delta r^{2}, \hspace{10.mm} \mu_{1}=r\mu_{2}=r_{12}p_{2}=\delta r r_{12}.\]  
These coefficients satisfy
\[c^{2}(p_{1}+\mu_{2})=c^{2}(1+p_{2})=c_{0}^{2}+c_{1}^{2}, \hspace{10.mm}
c^{4}(p_{1}\mu_{2}-p_{2}\mu_{1})=c^{4}\mu_{2}=c_{0}^{2}c_{1}^{2}\]
where the intrinsic barycentric and baroclinic phase speeds are given by
\[c_{0/1}^{2}=\frac{1}{2}c^{2}(1+\delta r \pm \sqrt{(1-\delta r)^{2}+4\delta r^{2}}).\]
Taking the curl of (3) and (4) one arrives at the perturbation vorticity budgets
\begin{equation}
R^{2}d_{0}z+j^{a}\partial_{a}\,F_{0}=-W^{n}\partial_{n}\,R^{2}\zeta-2i^{a}\partial_{a}\,S
\end{equation}
\begin{equation}
R^{2}d_{1}\zeta+i^{a}\partial_{a}\,F_{1}=-r_{12}(W^{n}\partial_{n}\,R^{2}z+
2j^{a}\partial_{a}\,S)
\end{equation}
with barycentric perturbation vorticity
\begin{equation}
R^{2}z=\epsilon^{an}j_{n};_{a}-F_{0}m-2SR\eta
\end{equation}
and baroclinic perturbation vorticity
\begin{equation}
R^{2}\zeta=\epsilon^{an}i_{n};_{a}-F_{1}R\eta-2r_{12}Sm.
\end{equation}
Equations (1) through (8) pose the Rayleigh stability problem for a generic stably stratified
and vertically sheared fluid on the rotating spherical surface. In spherical bishallow water
theory, constant mean layer-thicknesses exclude a mean flow with available potential
energy. The tidal problem for a circulation with finite available potential energy is 
discussed in the Appendix.

\vspace{5.mm}
\noindent
{\bf III. Quasigeostrophic Stability Analysis}

In the barotropic 1-layer limit, equations (1) through (4) reduce to Laplace's standard
tidal equations. In special cases, exact analytical solutions of the tidal equation are 
known in terms of confluent Heun functions, namely spheroidal functions \cite{7,8}. In the 
entire wave number space, approximate analytical solutions can be expressed in terms of 
spheroidal functions and the asymptotics of tidal functions coincide with the asymptotic 
behaviour of prolate spheroidal functions: the Margules regime of globally defined Legendre 
polynomials at small Lamb parameters \cite{9} and the Matsuno regime of Hermite polynomials 
on the equatorial $\beta$-plane at large Lamb parameters \cite{10}. Longuet-Higgins' 
quasigeostrophic Rossby wave filter \cite{4} retains these functional characteristics. In 
the 1-layer limit, the perturbation vorticity budget becomes
\[R^{2}d_{t}z+j^{a}\partial_{a}\,F=0\]
with perturbation vorticity
\begin{equation}
R^{2}z=\epsilon^{an}j_{n};_{a}-Fm.
\end{equation}
As in the strictly nondivergent case, quasigeostrophy assumes that the perturbation mass 
flux is sufficiently represented by a stream function $A$
\[j_{n}=\epsilon_{nm}\partial^{m}\,A\]
while the mass perturbation in (9) is not supposed to vanish, thus giving rise to the notion of
weakly divergent perturbations. A closed expression for (9) is now obtained by invoking the
geostrophic approximation
\[c^{2}m=-FA\]
leading for the vorticity budget to the equation
\[(\Delta-\alpha^{2}y^{2}-M\tau)A=0\]
where $y=sin\varphi$, $\alpha=2a(\Omega+U)/c$ the Lamb parameter, $M$ the zonal wave number, 
$\nu=a(\omega-UM)/c$ the Doppler-shifted frequency and $\tau=\alpha/\nu$. This is the prolate 
spheroidal wave equation. The dispersion relation for quasigeostrophic Rossby waves becomes
\[\nu=-\alpha M/\epsilon(N,M;\alpha)\]
with prolate spheroidal eigenvalue $\epsilon(N,M;\alpha)$. Comparison of this expression with
numerical solutions of the complete tidal equation (fig.1) demonstrates that with the 
exception (of the gravity branch) of the Yanai wave (mode number N=0) quasigeostrophy provides 
a satisfactory approximation to all
Rossby modes of the tidal problem. Also, this shows that quasigeostrophy is by no means a
regional, e.g.\ extratropical approximation. Rather, it is globally valid and includes the 
Margules regime as well as the Matsuno regime. Physically,  the equatorial $\beta$-plane 
approximation of the prolate spheroidal equation accounts for wave trapping in the Yoshida 
guide. A similar wave guide in midlatitudes does not exist and a midlatitude $\beta$-plane 
does not appear in the systematic approximation theory of spheroidal functions \cite{8}.

\begin{figure}[!h]
\centerline{\hbox{
\psfig{figure=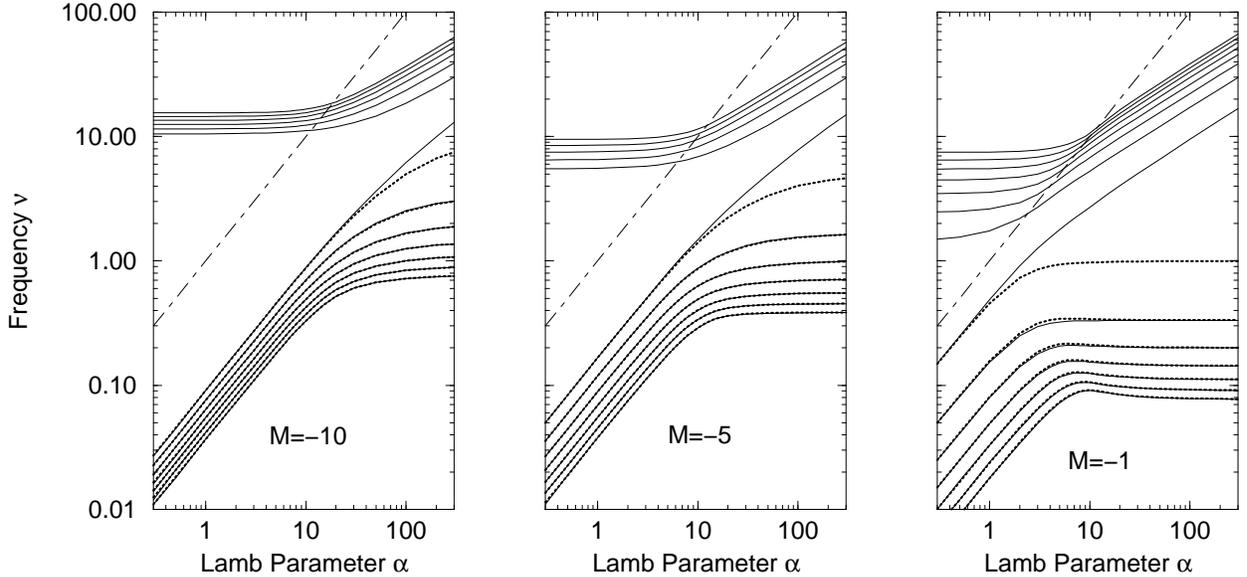,height=17.cm,angle=270,clip=}
}}
\caption{\small Tidal eigenfrequencies (solid lines) and quasigeostrophic 
approximation (dotted lines). Frequencies larger than $|M|$ correspond to gravity modes. 
Negative zonal wave numbers $M$ indicate westward propagation. Dashed-dotted line: $\nu=\alpha$. 
Tidal frequencies were calculated by [11] and spheroidal eigenvalues computed with NAG-Lib 
routine F02GJE.}
\end{figure}

The functional structure of the tidal problem uniquely determines the physical interpretation
of the so-called ``beta-effect'' and the spectrum of the tidal wave operator. The Coriolis
term of tidal theory represents the meridional shear of a mean zonal flow with uniform angular
velocity $\Omega+U$. Doppler-shifts only appear with respect to $U$ since the observer
corotates with $\Omega$. While the corotating observer does not see a frequency shift with
respect to $\Omega$, the corresponding meridional shear does not vanish on this 
transformation. Physically, the ``beta-effect'' of tidal theory refers to such meridional 
shear and differs profoundly from topography on the f-plane. 

The spectrum of the tidal wave operator represents free waves in an elastic medium. Such a 
medium has two types of excitations: longitudinal P (primary, pressure or sound) waves and 
transversal S (secondary or shear) waves. Long gravity waves of covariant shallow water 
theory are represented as longitudinal sound waves with restoring vertical buoyancy forces 
appearing as an effective ``compressibility'' of the strictly 2-dimensional system. 
Correspondingly, shallow baroclinic gravity waves are represented as second sound in
a strictly 2-dimensional bifluid. Rossby waves, on the other hand, obey the dynamics of
(radially polarized) shear waves \cite{12}. The frequencies of S-waves are always lower than 
P-wave frequencies: all Rossby frequencies lie below gravity frequencies. Rossby waves are 
transversal and mean flow shear exerts restoring stresses tangential to wave fronts. 
Low-frequency Rossby waves are essentially divergence-free and well represented by Margules' 
approximation. High Rossby frequencies are limited by meridional trapping at low latitudes 
which induces a weak divergence: weak divergence is characteristic of equatorial Rossby 
waves. Tropically trapped, weakly divergent Rossby waves are well approximated by Matsuno's 
theory. Longuet-Higgens' quasigeostrophy unifies both approaches. With two Lam\'{e} 
coefficients, elastic wave theory is inherently a two-parameter problem. This is also true 
for generic tidal theory: while  a ``compressibility'' controls gravity wave dynamics, 
the Lamb parameter represents the mean meridional shear that governs Rossby wave dynamics. 
Quasigeostrophy is the shear wave filter of tidal theory.

In application to the present bishallow problem, quasigeostrophy represents the barycentric 
and baroclinic mass flux perturbations in terms of stream functions
\[j_{n}=\epsilon_{nm}\partial^{m}\,A, \hspace{10.mm} i_{n}=\epsilon_{nm}\partial^{m}\psi\]
and determines mass- and concentration-perturbations in (7) and (8) from the thermal 
wind relation of (3) and (4)
\[c^{2}p_{1}\,m+c^{2}p_{2}\,R\eta=-F_{0}A-2S\psi\]
\[c^{2}\mu_{1}\,m+c^{2}\mu_{2}\,R\eta=-F_{1}\psi-2r_{12}S A.\]
Solving for $m$ and $R\eta$ and inserting the result into (7) and (8) the perturbation 
vorticities become
\[R^{2}z=-(\Delta-h_{0}y^{2})A-hy^{2}\psi\]
\[R^{2}\zeta=-(\Delta-h_{1}y^{2})\psi-r_{12}hy^{2}A\]
with
\[h_{0}=(\mu_{2}\alpha_{(2)}^{2}+r_{12}\alpha_{12}^{2})/\mu_{2}, \hspace{10.mm} h_{1}=
(rp_{2}\alpha_{(2)}^{2}+\alpha_{1}^{2})/\mu_{2}, \hspace{10.mm} h=
(p_{2}\alpha_{(2)}^{2}-\alpha_{12}\alpha_{1})/\mu_{2}\]
where $\alpha_{(\ell)}=2a(\Omega+U_{(\ell)})/c$, $\alpha_{12}=\alpha_{(1)}-\alpha_{(2)}$ and
$\alpha_{1}=2a(\Omega+U_{1})/c$. With these expressions the vorticity budgets (5) and (6) 
assume the form
\begin{equation}
(\omega_{0}\Delta-H_{00}y^{2}-f_{0}M)A=(WM(\Delta+2)-H_{01}y^{2})\psi
\end{equation}
\begin{equation}
(\omega_{1}\Delta-H_{11}y^{2}-f_{1}M)\psi=r_{12}(WM(\Delta+2)-H_{10}y^{2})A.
\end{equation}
Here,  $\omega_{0/1}=\omega-U_{0/1}M$ and $f_{0/1}=2(\Omega+U_{0/1})$ while
\[H_{00}=\omega_{0}h_{0}+r_{12}hWM, \hspace{5.mm} H_{01}=\omega_{0}h+h_{1}WM\]
and 
\[H_{10}=\omega_{1}h+h_{0}WM, \hspace{5.mm} H_{11}=\omega_{1}h_{1}+r_{12}hWM. \]
Equations (10) and (11) are a system of coupled spheroidal equations and spheroidal functions 
form a complete set of eigensolutions. Eliminating $\Delta \psi$ from these equations yields
\[q\psi=(\Delta-h_{0}y^{2}-M\tau_{0})A\]
with $\tau_{(\ell)}=2(\Omega+U_{(\ell)})/(\omega-U_{(\ell)}M)$ and
\[q=M\tau_{12}-hy^{2}, \hspace{5.mm} \tau_{12}=\tau_{(1)}-\tau_{(2)}, \hspace{5.mm}
\tau_{0}=r\tau_{(1)}+(1-r)\tau_{(2)}, \hspace{5.mm}\tau_{1}=(1-r)\tau_{(1)}+r\tau_{(2)}.\]
Substituting this expression for the baroclinic stream function into (10) results in a single
fourth-order equation
\begin{equation}
(\Delta-h_{1}y^{2}-M\tau_{1})\frac{1}{q}(\Delta-h_{0}y^{2}-M\tau_{0})A=r_{12}qA
\end{equation} 
for the barycentric stream function. Utilizing now the spheroidal property: 
$\Delta A=(\beta^{2}y^{2}-\epsilon)A$ with
\[\epsilon+M\tau_{0}-(\beta^{2}-h_{0})y^{2}=Xq\]
for constant $X$, equation (12) requires the simultaneous validity of the two quadratic 
equations
\[\tau_{12}(X^{2}+(1-2r)X-r_{12})=0\]
\[X^{2}h+(h_{0}-h_{1})X-r_{12}h=0.\]
These two equations express the major difference between the cartesian and the spherical
stability problem. In cartesian geometry, the dispersion relation determines admissible
eigenfrequencies. On the rotating spherical surface, background inhomogenities due to the
planet's sphericity, coordinate-dependent Coriolis forces and the mean circulation also
determine admissible Lamb parameters. Rayleigh theory of spatially inhomogeneous systems
accounts for wave trapping. The compatibility of both quadratic equations is determined 
by the relation 
\begin{equation}
h_{0}-h_{1}=(1-2r)h-\alpha_{(1)}\alpha_{(2)}/\mu_{2}
\end{equation}
which follows from the definition of $h_{0}$, $h_{1}$ and $h$. Given (13), essentially two 
classes of solutions of (12) exist: either, the stratified background is free of vertical 
shear or Rossby waves are strictly nondivergent. 
In the first case: $W=0$ and $\tau_{12}=0$. Hence: $\tau_{0}=\tau_{1}=\tau$, while $q=-hy^{2}$ 
and (12) becomes
\[(\Delta-\beta_{1}^{2}y^{2}-M\tau)\frac{1}{y^{2}}(\Delta-\beta_{0}^{2}y^{2}-M\tau)A=0\]
where the Lamb parameters are obtained as
\[\beta_{0}=2a(\Omega+U)/c_{0}, \hspace{10.mm} \beta_{1}=2a(\Omega+U)/c_{1}\]
with intrinsic barycentric and baroclinic phase speeds $c_{0/1}$. The dispersion relation is
\[(\nu_{0}\,\epsilon(N,M;\beta_{0})+\beta_{0}M)(\nu_{1}\,\epsilon(N,M;\beta_{1})+\beta_{1}M)=0\]
with prolate spheroidal eigenvalue $\epsilon(N,M;\beta)$ and $\nu_{0/1}=a(\omega-UM)/c_{0/1}$. 
In the shear-free case, Rossby waves are weakly divergent, propagate as barycentric and 
baroclinic modes and eigenfrequencies are real. In the second case:\,$c\rightarrow\infty$ and 
$h_{0}=h_{1}=h=0$, while $q=M\tau_{12}=const$ so that (12) reduces to
\[(\Delta-M\tau_{(1)})(\Delta-M\tau_{(2)})A=0.\]
In the strictly divergence-free case, spheroidal functions degenerate into Legendre 
polynomials and the dispersion relation becomes
\[(\epsilon+M\tau_{(1)})(\epsilon+M\tau_{(2)})=0\]
with $\epsilon(N,M)=N(N+1)+(2N+1)|M|+M^{2}$. The mean vertical shear traps nondivergent 
Rossby waves in individual layers and eigenfrequencies are real. This remains true if 
the mean circulation exhibits available potential energy \cite{13}. 

There are two more solutions if in one of the layers $\alpha_{(\ell)}=0$, 
i.e.\ $U_{(\ell)}=-\Omega$. For $\alpha_{(j)}=0$, Rossby waves exist only in the 
complementary layer $\ell\neq j$ and from (12) one finds the dispersion relation
\[\epsilon(N,M;\beta_{(\ell)})=-M\tau_{(\ell)}\]
with Lamb parameter
\[\beta_{(\ell)}^{2}=4a^{2}(\Omega+U_{(\ell)})^{2}/g'H_{(\ell)}\]
where $H_{(\ell)}=R_{(\ell)}/\rho_{(\ell)}$ is the mean layer-thickness and $g'= 
g(\rho_{(2)}-\rho_{(1)})/\rho_{(2)}$ the reduced gravity accelaration. Both of these
solutions are weakly divergent and stable. The condition $U_{(j)}=-\Omega$ implies a
westward flow circulating the globe in one day. Hence, the layer is at rest in a
nonrotating, inertial system and such conditions do not admit Rossby wave propagation 
(the mean meridional shear is absent). In the rotating system, corresponding large-scale 
velocities are of a magnitude that is not met on this planet. For practical purposes 
these solutions are hence of little significance. 

The same dispersion relations follow if $A$ and $\psi$ in (10) and (11) are replaced with 
spheroidal functions and the resulting system of 3 algebraic equations is solved for the 
unknown Lamb parameters and eigenfrequencies (see Appendix).  As a further alternative, the 
layer representation of spherical linearized bishallow water \cite{13} may be chosen as 
starting point for the stability analysis rather than the modal representation (1) through 
(4). It is readily seen that the dispersion relation from such an approach coincides with 
the results obtained above (see Appendix). The baroclinic instability problem of tidal 
theory does not assume the form of a spherical Taylor-Couette flow and all eigenfrequencies 
are real. Unlike baroclinic gravity waves, Rossby waves do not feed on the energy of a 
stably stratified and vertically sheared mean flow. Thus, the flow is linearly stable 
against quasigeostrophic disturbances. 

For the system (1) through (4) isopycnals coincide with equipotential surfaces. This type
of configuration is generally considered in baroclinic instability theory: isopycnals are 
assumed to be ``flat'' and a slope-parameter does not enter the problem \cite{3}. On the 
other hand, observers and modellers are typically concerned with sloped isopycnals and their 
erosion by baroclinic instability \cite{2}. The tidal equations for bishallow water with 
sloping isopycnals are well known and their stability against nondivergent Rossby waves has 
been demonstrated \cite{13}. The stability of such a system against quasigeostrophic Rossby 
waves is shown in the Appendix.

The physical interpretation of these results is best considered in comparison to 
Kelvin-Helmholtz instability. For this instability, highly divergent baroclinic gravity
waves continuously sample both layers of a stably stratified and vertically sheared fluid.
The dynamics of baroclinic gravity waves are controlled by the competition of (stabilizing)
stratification and (destabilizing) vertical shear. If the vertical shear becomes too
large, baroclinic gravity waves grow. None of these mechanisms plays a role in tidal
Rossby wave dynamics. Rossby waves are inseparably linked to a definite value of the mean 
meridional shear determined by $\Omega+U_{(\ell)}$. Hence, they propagate in individual 
layers and a distinction of barycentric and baroclinic Rossby waves is meaningless (unless
the vertical shear vanishes). Weak divergencies are associated with meridional trapping
and do not accommodate the exploration of adjacent layers. None of the mean flow features
of a stably stratified, vertically sheared flow competes with the restoring stresses. Thus, 
a transfer of energy or vorticity between this type of circulation and Rossby waves is 
excluded in the framework of tidal theory. 

Although mean flow available potential energy may alter the system's wave guide geography
radically, it does not change its stability properties. For very low-frequency Rossby waves,  
isopycnal slopes modify the effective Lamb parameter (see 
Appendix). This parameter may now assume real or imaginary values. In the event of imaginary 
Lamb parameters, spheroidal wave operators change from prolate to oblate \cite{8} and Rossby 
waves may be meridionally trapped in a polar wave guide \cite{13,14,15}. Although the Lamb 
parameter of oblate Rossby waves becomes imaginary, their frequencies remain real \cite{8} 
and Rossby wave amplitudes do not commence growing. In general, a stably stratified and 
vertically sheared fluid with or without available potential energy lacks the faculties of 
energy- and vorticity-transfer to Rossby waves. This statement may require modification for 
certain initial conditions or the extremely steep isopycnals associated with outcropping. 
Independent of the role, these and other special cases may take, Laplace's tidal theory does 
not support the ubiquity, baroclinic instability is currently thought to have for irregular 
ocean-atmosphere motions.

\vspace{5.mm}
\noindent
{\bf IV. Discussion} 

The results of the previous section are in clear contrast to baroclinic instability theory 
on the midlatitude $\beta$-plane. To identify the source of this discrepancy evaluate the 
midlatitude $\beta$-plane approximation of (12). In this sense, the latitude $y$ is fixed at 
some (extratropical) value $y_{\ast}$ while $G_{0}^{2}=h_{0}y_{\ast}^{2}$, $G_{1}^{2}=
h_{1}y_{\ast}^{2}$ and $G^{2}=h y_{\ast}^{2}$ are considered as constant parameters. Adopting 
furthermore a cartesian Laplace operator and trigonometric eigenfunctions with $K^{2}=
a^{2}(k_{1}^{2}+k_{2}^{2})$ and $M=a\,k_{1}\,cos\varphi_{\ast}$ one obtains from (12) the 
approximate dispersion relation
\[(K^{2}+G_{1}^{2}+M\tau_{1})(K^{2}+G_{0}^{2}+M\tau_{0})=r_{12}q^{2}.\]
A little algebra readily shows that this expression is equivalent to
\[(K^{2}+F_{(1)}^{2}+M\tau_{(1)})(K^{2}+F_{(2)}^{2}+M\tau_{(2)})=s F_{(1)}^{2}F_{(2)}^{2}\]
with $s=\rho_{(1)}/\rho_{(2)}$ and $F_{(\ell)}^{2}=\beta_{(\ell)}^{2}y_{\ast}^{2}$. This 
equation is quadratic in frequency.
Unlike the results of the previous section, its roots do not assume an easily interpreted
form. This is indicative of the difficulties of the primarily geometrical midlatitude
$\beta$-plane approximation to accommodate the system's physical structure. Nevertheless, 
the reality of these roots is obvious and the midlatitude $\beta$-plane approximation to the 
final wave equation (12) comes qualitatively to the same result as the spherical analysis with 
respect to stability. Hence, the trigonometric approximation of wave functions in 
itself - though unsatisfactory - is uncritical. The source of discrepancies is therefore a 
fundamentally different account of vorticity fluxes by covariant shallow waters and baroclinic 
instability theory. 

In the covariant case, the vorticity budget (12) is uniquely determined by the equations 
of motion (1) through (4) and the quasigestrophic thermal wind approximation. The validity 
of quasigeostrophy for the barotropic fluid is well demonstrated by fig.1 and the particular 
form of the same argument for the baroclinic system is again an unambiguous consequence of 
the equations of motion. These equations are the direct and unique result of the application
of covariance requirements to the formulation of spherical shallow water dynamics. Thus, 
the vorticity fluxes of covariant shallow water theory are in essence the expression of basic
geometrical and physical consistency conditions for the hydrostatic flow of the stably 
stratified and vertically sheared fluid on the rotating spherical surface.

Standard baroclinic instability theory, on the other hand, takes the Primitive Equations 
\cite{16} as a starting point and invokes the midlatitude $\beta$-plane approximation to 
derive the vorticity budget \cite{3}. It has been shown that the Primitive Equations do 
not pose a covariant dynamical problem and involve ambiguous mass and momentum fluxes as 
a consequence of the violation of Newton's first law \cite{6}. For the vorticity budget 
it now becomes important that the midlatitude $\beta$-plane is much more poorly defined 
than the equatorial $\beta$-plane. Effectively, it takes the role of a geometric closure 
assumption which introduces ill-defined vorticity fluxes. In the stability analysis this 
inconsistency resurfaces as spurious Rossby wave growth, i.e. baroclinic instability.

Vorticity fluxes are crucial in maintaining and changing the density field of the
ocean-atmosphere system and a consistent representation of vorticity dynamics is 
indispensable for realistic numerical simulations of the global circulation. At this 
time, a large number of numerical circulation models is based on the Primitive Equations. 
Moreover, most models are not formulated in terms of spherical coordinates but utilize a 
multi-$\beta$-plane approach to approximate the globe's sphericity: Laplace operators, for 
instance, are coded as a sum of second order derivatives, ignoring first order contributions 
from nontrivial Christoffel symbols. Widely independent of spatio-temporal resolution and the 
quality of subscale parametrizations, such models cannot expect to simulate the large-scale 
circulation with geometric-dynamic integrity and realism. 

Theoretically as well as numerically, large-scale ocean-atmosphere dynamics require a 
covariant dynamical framework including the acknowledgement of the spherical geometry of the 
planet's surface. The linear stability of the baroclinic fluid against quasigeostrophic 
disturbances calls for a covariant reanalysis of the observational evidence on dynamical 
instabilities in the ocean-atmosphere system.

\vspace{5.mm}
\noindent
{\bf Appendix}

The effect of available potential energy of the circulation is considered. With constant
surface pressure, the mean layer mass per unit area is now assumed to vary with latitude 
according to 
\[R_{(\ell)}=R_{E}^{(\ell)}(1-b_{(\ell)}y^{2})\] 
where $R_{E}^{(\ell)}$ is its equatorial and $R_{P}^{(\ell)}$ its polar value. As a 
geostrophic solution of the nonlinear bishallow water equations, its slope parameters  
\[b_{(\ell)}=(R_{E}^{(\ell)}-R_{P}^{(\ell)})/R_{E}^{(\ell)}\]
satisfy 
\[b_{(1)}=a^{2}(2\Omega+U_{(1)}+U_{(2)})W/2\gamma R_{E}^{(1)}\approx
 a^{2}\Omega W/\gamma R_{E}^{(1)}.\]
\[b_{(2)}=a^{2}[(2\Omega+U_{(2)})(\delta U_{(2)}-W)-U_{(1)}W]/2\gamma R_{E}^{(2)}\approx
a^{2}\Omega (\delta U_{(2)}-W)/\gamma R_{E}^{(2)}.\]
with $\gamma=\gamma_{(2)}\delta$. The associated mean vertical shear $W=U_{(1)}-U_{(2)}$ is 
in thermal wind balance and the Coriolis parameter in layer $\ell$ is given by 
\[F_{(\ell)}=R_{(\ell)}Z_{(\ell)}=2(\Omega+U_{(\ell)})sin\varphi.\] 
With $d_{(\ell)}=\partial_{t}+U_{(\ell)}\partial_{\lambda}$, the
layer representation of this tidal problem assumes the form \cite{13}
\[d_{(\ell)}m_{(\ell)}+j^{n}_{(\ell)};_{n}=0\]
\[d_{(\ell)}v_{n}^{(\ell)}+\epsilon_{mn}F_{(\ell)}v^{m}_{(\ell)}+\partial_{n}p_{(\ell)}=0\]
where the perturbation pressures are given by
\[p_{(1)}=\gamma_{(1)}m_{(1)}+\gamma_{(2)}m_{(2)}, \hspace{5.mm}
p_{(2)}=\gamma_{(2)}(m_{(1)}+m_{(2)})=\gamma_{(2)}m.\]
The curl of the momentum budgets yields the vorticity budgets
\[R_{(\ell)}d_{(\ell)}z_{(\ell)}+j^{a}_{(\ell)}\partial_{a}Z_{(\ell)}=0\]
for the perturbation vorticities
\[R_{(\ell)}z_{(\ell)}=\epsilon^{an}v_{n}^{(\ell)};_{a}-Z_{(\ell)}m_{(\ell)}.\]
The quasigeostrophic approximation with
\[j_{n}^{(\ell)}=\epsilon_{nm}\partial^{m}A_{(\ell)}\]
and the thermal wind relation
\[\gamma m_{(1)}=Z_{(2)}A_{(2)}-Z_{(1)}A_{(1)}, \hspace{5.mm}
s\gamma m_{(2)}=sZ_{(1)}A_{(1)}-Z_{(2)}A_{(2)}\]
leads to the coupled wave equations
\[\gamma[(R_{(1)}\Delta-R_{(1)}^{a}\partial_{a}-F_{(1)}^{2}/\gamma)d_{(1)}-
\epsilon^{an}R_{(1)}^{2}\partial_{a}Z_{(1)}\partial_{n}]A_{(1)}=
-R_{(1)}Z_{(2)}F_{(1)}d_{(1)}A_{(2)}\]
\[\gamma[(R_{(2)}\Delta-R_{(2)}^{a}\partial_{a}-F_{(2)}^{2}/s\gamma)d_{(2)}-
\epsilon^{an}R_{(2)}^{2}\partial_{a}Z_{(2)}\partial_{n}]A_{(2)}=
-R_{(2)}Z_{(1)}F_{(2)}d_{(2)}A_{(1)}.\]
For $0\leq |b_{(\ell)}| \ll 1$ these equations are a coupled system of spheroidal equations.
Physically, this condition on the slope parameters excludes outcropping of isopycnals. With
\[\beta_{(\ell)}^{2}=4a^{2}(\Omega+U_{(\ell)})^{2}/g'H_{E}^{(\ell)}, \hspace{5.mm}  
h_{(\ell)}=\beta_{(\ell)}^{2}+2b_{(\ell)}M\tau_{(\ell)}\]
the quasigeostrophic vorticity budgets assume the form
\[(\Delta-h_{(1)}y^{2}-M\tau_{(1)})A_{(1)}=-sk\beta_{(1)}\beta_{(2)}y^{2}A_{(2)}\]
\[k(\Delta-h_{(2)}y^{2}-M\tau_{(2)})A_{(2)}=-\beta_{(1)}\beta_{(2)}y^{2}A_{(1)}\]
where $k=\sqrt{H_{E}^{(1)}/H_{E}^{(2)}}$. Note that $h_{(\ell)}$ will become negative for
very low Rossby wave frequencies at negative $b_{(\ell)}M$. Using now 
$\Delta A_{(\ell)}=(\beta^{2}y^{2}-\epsilon)A_{(\ell)}$, these equations reduce to a system 
of coupled algebraic equations for the amplitudes of the stream functions. Nontrivial 
solutions exist for vanishing coefficients of the polynomial $X+Yy^{2}+Zy^{4}$. For the 
present problem, these coefficients have the form
\[X=(\epsilon+M\tau_{(1)})(\epsilon+M\tau_{(2)})=0\]
\[Y=(\epsilon+M\tau_{(1)})(\beta^{2}-h_{(2)})+(\epsilon+M\tau_{(2)})(\beta^{2}-h_{(1)})=0\]
\[Z=\beta^{4}-(h_{(1)}+h_{(2)})\beta^{2}+h_{(1)}h_{(2)}-s\beta_{(1)}^{2}\beta_{(2)}^{2}=0\]
The first of these equations states the reality of eigenfrequencies for stream functions
which remain regular at the poles. Also, it demonstrates that Rossby waves propagate in
layers unless the vertical shear vanishes. The second equation selects the Lamb parameter
for the respective mode and the third equation determines the Lamb parameters. For 
$b_{(\ell)}=0$ the solutions of the main text emerge while the case $c\rightarrow\infty$ 
and $b_{(\ell)}\neq 0$ has been discussed in \cite{13}.


\begin{thebibliography}{99}
\bibitem{1}W.Munk, in: 50 Years of Ocean Discovery, 44 (National Academy Press, 
Washington, 2000).
\bibitem{2}S.Griffies et al., Ocean Modelling {\bf 2}, 123 (2000).
\bibitem{3}J.Pedlosky, {\em Geophysical Fluid Dynamics} (Springer, New York, 1983).
\bibitem{4}M.Longuet-Higgins, Proc.R.Soc.London, Ser.A {\bf 284}, 40 (1965).
\bibitem{5}R.Pierrehumbert and K.Swanson, Ann.Rev.Fluid Mech. {\bf 27}, 419 (1995).
\bibitem{6}D. M\"{u}ller, Int.J.Mod.Phys.B {\bf 11}, 223 (1997).
\bibitem{7}A.Ronveaux, Ed., {\em Heun's Differential Equations} (Oxford, New York, 1995).
\bibitem{8}C.Flammer, {\em Spheroidal Wave Functions} (Stanford University Press, 
Stanford, 1957).
\bibitem{9}M.Margules, Sitzungsber.Kais.Akad.Wiss.Wien Math.Naturwiss.{\bf 102}, 11 (1893).
\bibitem{10}T.Matsuno, J.Meteorol.Soc.Jpn.{\bf 44}, 23 (1966).
\bibitem{11}P.Swarztrauber and A.Kasahara, SIAM (Soc.Ind.Appl.Math.) J.Sci.Stat.Comput.{\bf 6}, 
464 (1985).
\bibitem{12}J.Achenbach, {\em Wave Propagation in Elastic Solids} (North Holland, New York, 1973).
\bibitem{13}D. M\"{u}ller and E.Maier-Reimer, Phys.Rev.E {\bf 61}, 1468 (2000).
\bibitem{14}M.Longuet-Higgins, Philos.Trans.R.Soc.London, Ser.A {\bf 262}, 511 (1968).
\bibitem{15}S.Chapman and R.Lindzen, {\em Atmospheric Tides} (Reidel, Dordrecht, 1970).
\bibitem{16}L.Richardson, {\em Weather Prediction by Numerical Process} (Cambridge University
Press, Cambridge, 1922). 
\end{thebibliography}
\end{document}